# METHODS FOR EVALUATING SOFTWARE ACCESSIBILITY

**Kuz Mykola** – Dr. Sc., Professor of the Department of Information Technology Vasyl Stefanyk Precarpathian National University, Ivano-Frankivsk, Ukraine.

**Yaremiy Ivan** – Dr. Sc., Professor of the Department of Applied Physics and Materials Science, Vasyl Stefanyk Precarpathian National University, Ivano-Frankivsk, Ukraine.

**Yaremii Hanna** – MSc., Vasyl Stefanyk Precarpathian National University, Ivano-Frankivsk, Ukraine.

**Pikuliak Mykola** – PhD, Associate Professor of the Department of Information Technology Vasyl Stefanyk Precarpathian National University, Ivano-Frankivsk, Ukraine.

**Lazarovych Ihor** – PhD, Associate Professor of the Department of Information Technology Vasyl Stefanyk Precarpathian National University, Ivano-Frankivsk, Ukraine.

**Kozlenko Mykola** – PhD, Associate Professor of the Department of Information Technology Vasyl Stefanyk Precarpathian National University, Ivano-Frankivsk, Ukraine.

**Vekeryk Denys** – MSc., Vasyl Stefanyk Precarpathian National University, Ivano-Frankivsk, Ukraine.

## ABSTRACT

**Context**. The development and enhancement of methods for evaluating software accessibility is a relevant challenge in modern software engineering, as ensuring equal access to digital services is a key factor in improving their efficiency and inclusivity. The increasing digitalization of society necessitates the creation of software that complies with international accessibility standards such as ISO/IEC 25023 and WCAG. Adhering to these standards helps eliminate barriers to software use for individuals with diverse physical, sensory, and cognitive needs. Despite advancements in regulatory frameworks, existing accessibility evaluation methodologies are often generalized and fail to account for the specific needs of different user categories or the unique ways they interact with digital systems. This highlights the need for the development of new, more detailed methods for defining metrics that influence the quality of user interaction with software products.

**Objective**. Building a classification and mathematical model and developing accessibility assessment methods for software based on it.

**Methods**. A method for assessing the quality subcharacteristic "Accessibility", which is part of the "Usability" quality characteristic, has been developed. This enabled the analysis of a website's inclusivity for individuals with visual impairments, and the formulation of specific recommendations for further improvements, which is a crucial step toward creating an inclusive digital environment.

**Results**. Comparing to standardized approaches, a more detailed and practically oriented accessibility assessment methodology has been proposed. Using this methodology, an analysis of the accessibility of the main pages of Vasyl Stefanyk Precarpathian National University's website was conducted, and improvements were suggested to enhance its inclusivity.

**Conclusions.** This study presents the development of a classification and mathematical model, along with an accessibility assessment methodology for websites based on the ISO 25023 standard, and an analysis of the main pages of the university's web portal. The identified quantitative accessibility indicators enable an evaluation of the web resource's compliance with modern inclusivity requirements and provide recommendations for its improvement.

The scientific novelty of this research lies in the development of assessment methods for the "Accessibility" quality subcharacteristic by introducing new subproperties and attributes of software quality, based on clearly defined metrics specifically adapted for evaluating the accessibility level of digital products for individuals with visual impairments. This approach ensures a more precise and objective determination of web resources' compliance with inclusivity requirements, contributing to their effectiveness and usability for this user group.

The practical significance of the obtained results lies in their applicability for objectively evaluating the accessibility of software products and web resources.

## ABBREVIATIONS

ISO is an International Organization for Standardization;
IEC is an International Electrotechnical Commission;
DSTU is a National Standard of Ukraine;
WCAG is a Web Content Accessibility Guidelines.

## NOMENCLATURE

$X_{UAC-1-G}$ is an "Accessibility for users with disabilities" quality property;

$A_{UAC-1-G}$ is a number of functions successfully used by the users with a specific disability;

$B_{UAC-1-G}$ is a number of functions implemented;

$X_{UAC-2-S}$ is a "Supported languages adequacy" quality property;

$A_{UAC-2-S}$ is a number of languages actually supported;

$B_{UAC-2-S}$ is a number of languages needed to be supported;

$X_{UAC-1.1.1-G}$ is an "Alternative text" quality attribute;

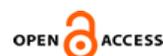
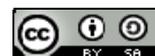

$A_{UAC-1.1.1-G}$ is a number of multimedia elements with meaningful text alternatives;

$B_{UAC-1.1.1-G}$ is a total number of multimedia elements in the system;

$X_{UAC-1.1.2-G}$ is a "Color contrast" quality attribute;

$A_{UAC-1.1.2-G}$ is a contrast level;

$B_{UAC-1.1.2-G}$ is a total number of elements;

$B_{UAC-1.1.2-G^+}$ is a number of elements that meet the following conditions: contrast level ≥ 4.5:1 for main text and ≥ 3:1 for auxiliary text;

$X_{UAC-1.1.3-G}$ is a "Subtitles and audio descriptions" quality attribute;

$A_{UAC-1.1.3-G}$ is a number of videos with subtitles or audio descriptions;

$B_{UAC-1.1.3-G}$ is a total number of videos in the system;

$X_{UAC-1.2.1-G}$ is a "Keyboard navigation" quality attribute;

$A_{UAC-1.2.1-G}$ is a number of interactive elements accessible via keyboard navigation;

$B_{UAC-1.2.1-G}$ is a total number of interactive elements in the system;

$X_{UAC-1.2.2-G}$ is a "Structured navigation" quality attribute;

$A_{UAC-1.2.2-G}$ represents presence or absence of breadcrumbs (1 if present, 0 if absent);

$B_{UAC-1.2.2-G}$ is a number of skipped heading levels;

$C_{UAC-1.2.2-G}$ is a total number of titles;

$X_{UAC-1.3.1-G}$ is a "Clear instructions" quality attribute;

$A_{UAC-1.3.1-G}$ is a number of instructions rated as clear;

$B_{UAC-1.3.1-G}$ is a total number of instructions;

$X_{UAC-1.3.2-G}$ is a "Input assistance" quality attribute;

$A_{UAC-1.3.2-G}$ is a number of fields with autocomplete or hint functions;

$B_{UAC-1.3.2-G}$ is a total number of fields that could have autocomplete or hint functions;

$X_{UAC-1.3.3-G}$ is a "Correct input support" quality attribute;

$A_{UAC-1.3.3-G}$ is a number of forms with error messages;

$B_{UAC-1.3.3-G}$ is a total number of forms in the system;

$X_{UAC-1.1-G}$ is a "Perceptiveness" quality subproperty;

$w_{UAC-1.1.1-G}$ is a weight of "Alternative text" quality attribute;

$w_{UAC-1.1.2-G}$ is a weight of "Color contrast" quality attribute;

$w_{UAC-1.1.3-G}$ is a weight of "Subtitles or audio descriptions" quality attribute;

$X_{UAC-1.2-G}$ is an "Operability" quality subproperty;

$w_{UAC-1.2.1-G}$ is a weight of "Keyboard navigation" quality attribute;

$w_{UAC-1.2.2-G}$ is a weight of "Structured navigation" quality attribute;

$X_{UAC-1.3-G}$ is a "Understandability" quality subproperty;

$w_{UAC-1.3.1-G}$ is a weight of "Clear instructions" quality attribute;

$w_{UAC-1.3.2-G}$ is a weight of "Input assistance" quality attribute;

$w_{UAC-1.3.3-G}$ is a weight of "Correct input support" quality attribute;

$X_{UAC-2.1-S}$ is a "Localization" quality subproperty;

$w_1$ is weight of state language;

$A_{UAC-2.1-S}$ represents presence of the state language;

$w_2$ is a weight of English language;

$B_{UAC-2.1-S}$ represents presence of the English language;

$w_3$ is a weight of popular European languages;

$C_{UAC-2.1-S}$ represents presence of one of the popular European languages

$w_4$ is weight of other languages;

$D_{UAC-2.1-S}$ represents presence of other languages;

$w_{UAC-1.1-G}$ is a weight of "Perceptiveness" quality subproperty;

$w_{UAC-1.2-G}$ is a weight of "Operability" quality subproperty;

$w_{UAC-1.3-G}$ is a weight of "Understandability" quality subproperty;

$w_{UAC-2.1-S}$ is a weight of "Localization" quality subproperty;

$w_{UAC-1-G}$ is a weight of "Accessibility for users with disabilities" quality property;

$w_{UAC-2-S}$ is a weight of "Supported languages adequacy" quality property.

## INTRODUCTION

The development and assessment of software products in today's environment require consideration of a wide range of characteristics that define their quality. Among these, the subcharacteristic "Accessibility" is one of the most critical, as it determines how easily a software system can be used by the broadest possible range of users.

Accessibility is particularly relevant in the context of increasing focus on inclusivity, as ensuring that software products can be used by individuals with diverse physical, sensory, and cognitive needs not only promotes social equity but also expands the potential user base [1].

According to the ISO 25023 standard [2], the assessment of software product quality is based on metrics that quantitatively reflect their efficiency, usability, and accessibility.

The accessibility criterion for users with disabilities determines how successfully individuals with specific physical, cognitive, or sensory limitations can complete tasks within a system. The level of accessibility is measured as the ratio of successfully utilized functions to the total number of implemented functions. This approach provides an objective assessment of a system's usability for users with various disabilities. At the same time, applying this criterion highlights the need to consider the diverse requirements of individuals with different types of impairments, as their needs can vary significantly. For instance, users with sensory impairments may require enhanced visual elements, while those with physical disabilities may prioritize adaptive system functionalities to improve interaction.

The system's compliance with users' language support needs is assessed by evaluating the level of language support, which is determined as the proportion of actually supported languages relative to those required for effective use. The language support metric is crucial for evaluating the internationalization of a software product, particularly in multilingual regions or among users with diverse linguistic preferences. It helps determine how well the system accommodates linguistic diversity and whether it can provide a seamless user experience in a multilingual environment.

Although these metrics provide fundamental guidelines for assessing system quality, their application often encounters limitations due to their generalized nature. They do not always account for the specific use cases or the unique operating conditions of the software. This highlights the need for their expansion and refinement to better address users' specific needs and usage contexts. Such an approach would enable a more objective and effective accessibility evaluation, ultimately improving the overall quality of the software product.

The object of this study is the process of evaluating software accessibility.

The subject of the study is the methods for determining the qualitative assessment of accessibility.

The aim of this work is to develop a classification and mathematical model and, based on it, design methods for evaluating software accessibility.

## 1 PROBLEM STATEMENT

The problem of modeling software accessibility indicators for individuals with visual impairments is considered.

Let the subcharacteristic "Accessibility", in accordance with the ISO 25023 standard [2], include the quality attribute "Accessibility for users with disabilities" (UAC-1-G), which defines the extent to which potential users, including those with impairments or disabilities, can successfully use a software system or product. This quality attribute ensures that users with visual, auditory, motor, or cognitive impairments can complete their tasks within the system, potentially with the aid of assistive technologies. As specified in [2], this quality attribute is determined using the following formula:

$$X_{UAC-1-G} = \frac{A_{UAC-1-G}}{B_{UAC-1-G}}. \quad (1)$$

However, the methodology for determining parameter $A_{UAC-1-G}$ is not provided in standard [2].

The second component of the "Accessibility" subcharacteristic is the quality attribute "Supported languages adequacy" (UAC-2-S). This attribute evaluates the issue that users often encounter operational errors when attempting to use software in a language different from their native one. Misinterpretation of descriptions and messages leads to a decrease in accessibility. According to [2], this quality attribute is determined using the following formula:

$$X_{UAC-2-S} = \frac{A_{UAC-2-S}}{B_{UAC-2-S}}. \quad (2)$$

The methodology for determining parameter $B_{UAC-2-S}$ is not provided in standard [2] as well.

The following tasks are to be addressed: 1) the development of a classification and mathematical model of software accessibility indicators by introducing a new group of indicators – quality attributes – based on which the parameters $A_{UAC-1-G}$ and $B_{UAC-2-S}$ can be determined; 2) the development of methods for qualitative assessment of accessibility; 3) the validation of the proposed methodology using the example of a university web portal.

## 2 REVIEW OF THE LITERATURE

This study focuses on users with visual impairments, as they represent the most active group among individuals with disabilities using the internet [3]. The further analysis centers on adapting the ISO 25023 standard [2] for accessibility evaluation and expanding its metrics to better address the needs of this user category.

Currently, software quality assessment according to ISO 25010 [4] is conducted based on the quality properties defined in ISO 25023 [2]. This process involves evaluating quality characteristics and subcharacteristics, which collectively contribute to the overall assessment of a software product's quality. Specifically, ISO 25023 [2] provides a set of measures that enable the quantitative evaluation of software quality. The application of metric analysis methods in this process allows for the calculation of numerical metric values that describe the degree to which a software product meets defined requirements. According to ISO 24765 [5], a metric is defined as a numerical measure of the extent to which a product possesses a specific property, making it a crucial tool in the quality assurance process.

The existing standardized criteria for evaluating quality properties, as outlined in [2], lack a sufficient methodological framework for assessment. Specifically, standard [2] provides formulas for determining various quality properties and descriptions of formula parameters but does not include clear methodologies for their determination. This also applies to the "Accessibility" subcharacteristic. Additionally, the document does not define any criteria for establishing weighting coefficients for quality metrics.

Study [6] initiated research aimed at improving standardized methodologies for software quality assessment. The applied aspects of software quality assessment methods are presented in studies [7–9]. These works provide examples of the practical validation of software quality assessment methodologies and qualitative evaluation of a web forum, thereby demonstrating the effectiveness of the developed approaches.

An essential component of the methodology for evaluating software quality measures is the determination of weighting coefficients. One such method is presented in study [10]. It is based on expert evaluation of quality measures and includes tools for verifying the accuracy and reliability of expert decisions. The method described in [10] served as the foundation for the weighting coefficient determination methodology (significance levels) for software quality metrics, including characteristics, subcharacteristics, and quality attributes.

Based on the analysis of existing software quality evaluation methods, it was determined that, despite the availability of a developed methodology for assessing software quality, including methods for determining weighting coefficients, the primary drawback is the absence of methods for defining specific quality attributes, which represent the lowest level of software product quality metrics.

As a result, this work presents the development of a method for evaluating one of the quality subcharacteristics, "Accessibility", along with its constituent quality attributes: "Accessibility for users with disabilities" and "Supported languages adequacy".

### 3 MATERIALS AND METHODS

An effective accessibility assessment will ensure the inclusivity of the system, making it intuitive and functional for the widest possible range of users.

By breaking down the quality attributes of the "Accessibility" subcharacteristic into subproperties, four key subproperties have been identified:
– perceptiveness (UAC-1.1-G);
– operability (UAC-1.2-G);
– understandability (UAC-1.3-G);
– localization (UAC-2.1-S).

This approach enables a more detailed examination of various aspects of accessibility that directly impact the ability of users with disabilities to successfully interact with the system.

"Perceptiveness" is critical for ensuring accessibility, as users with visual or hearing impairments rely on text alternatives, color contrast, subtitles, and audio descriptions to effectively perceive information.

"Operability" defines how well the interface can be used with different input methods, including keyboard, mouse, and assistive devices, which is especially important for users with motor impairments.

"Understandability" is fundamental to creating an inclusive experience, as it ensures clarity of instructions, accessibility of error correction features, and support for proper data input, reducing cognitive load.

"Localization" ensures the adaptation of textual content and interface elements to the linguistic and cultural characteristics of users, enabling diverse audiences to interact effectively with the web resource. This is especially relevant in multilingual environments, where accurate translation and proper content structure play a key role in information perception and comprehension.

These quality subproperties were identified as priorities due to their direct impact on users' ability to successfully complete tasks within the system. They reflect the fundamental principles of accessibility outlined in standards such as WCAG [11] and ISO 25023 [2], allowing for a more detailed examination of aspects that pose the greatest challenges for users with specific limitations.

Let's take a closer look at each subproperty, starting with "Perceptiveness", one of the most critical aspects influencing how users perceive information.

The first critically important quality attribute is "Alternative text" (UAC-1.1.1-G), which defines the proportion of images that have correct textual descriptions. The formula for calculating this indicator is as follows:

$$X_{UAC-1.1.1-G} = \frac{A_{UAC-1.1.1-G}}{B_{UAC-1.1.1-G}}. \quad (3)$$

The second quality attribute is "Color contrast" (UAC-1.1.2-G), which evaluates the compliance of text-to-background contrast levels with established requirements. The formula for calculating this indicator is as follows:

$$X_{UAC-1.1.2-G} = \frac{\sum_{i=1}^{n}\left(A_{UAC-1.1.2-Gi} \cdot B_{UAC-1.1.2-Gi^+}\right)}{\sum_{i=1}^{n}\left(A_{UAC-1.1.2-Gi} \cdot B_{UAC-1.1.2-Gi}\right)}. \quad (4)$$

The final parameter of perceptiveness is "Subtitles and audio descriptions" (UAC-1.1.3-G), which assesses the proportion of video content that includes subtitles or audio descriptions, focusing on the accessibility of multime-dia content for users with visual impairments. This parameter is determined using the following formula:

$$X_{UAC-1.1.3-G} = \frac{A_{UAC-1.1.3-G}}{B_{UAC-1.1.3-G}}. \quad (5)$$

The next accessibility subproperty is "Operability" (UAC-1.2-G), which evaluates how easily users can interact with the interface using different input methods, such as a keyboard or mouse. This subproperty is critical for ensuring accessibility for users with visual impairments, who often rely on the keyboard as their primary navigation tool.

To evaluate the proportion of interactive elements that are accessible via keyboard navigation, the quality attribute "Keyboard navigation" (UAC-1.2.1-G) is introduced. This attribute is calculated using the following formula:

$$X_{UAC-1.2.1-G} = \frac{A_{UAC-1.2.1-G}}{B_{UAC-1.2.1-G}}. \quad (6)$$

The second quality attribute is "Structured navigation" (UAC-1.2.2-G). This parameter evaluates the proportion of pages with well-structured navigation, such as breadcrumbs, and a clear heading hierarchy, and is described by one of the following formulas:

$$X_{UAC-1.2.2-G} = 0.5 \cdot A_{UAC-1.2.2-G} + 0.5 \cdot \left(1 - \frac{B_{UAC-1.2.2-G}}{C_{UAC-1.2.2-G}}\right), \quad (7)$$

or

$$X_{UAC-1.2.2-G} = 1 - \frac{B_{UAC-1.2.2-G}}{C_{UAC-1.2.2-G}}. \quad (8)$$

The variable A takes a value of 1 or 0, depending on the presence or absence of breadcrumbs, respectively. Formula (7) is used if the page is located at a deep hierarchical level; otherwise, Formula (8) is applied.

The third identified subproperty of "Accessibility" is "Understandability", which evaluates how easily users can comprehend information and interface elements. This subproperty is critical for ensuring intuitive interaction with a website, particularly for users with visual impairments or cognitive disabilities. A clear and well-structured interface reduces errors, improves system usability, and enhances the overall user experience for a diverse audience.

To evaluate the clarity of form-filling instructions, we define the quality attribute "Clear Instructions" (UAC-1.3.1-G). This attribute analyzes whether users are provided with clear, specific, and understandable guidelines for completing forms. It is determined using the following formula:

$$X_{UAC-1.3.1-G} = \frac{A_{UAC-1.3.1-G}}{B_{UAC-1.3.1-G}}. \quad (9)$$

The next "Understandability" attribute is "Input Assistance" (UAC-1.3.2-G), which evaluates the proportion of forms that include features to prevent or correct input errors. These features include autofill, interactive hints, real-time data validation, and format input notifications. This quality attribute is designed to reduce errors and enhance user interaction with forms. It is calculated using the following formula:

$$X_{UAC-1.3.2-G} = \frac{A_{UAC-1.3.2-G}}{B_{UAC-1.3.2-G}}. \quad (10)$$

Another quality attribute of the "Understandability" subproperty is "Correct input support" (UAC-1.3.3-G), which evaluates the presence of functional error messages that help users identify and correct mistakes when entering data into forms. This attribute is determined using the following formula:

$$X_{UAC-1.3.3-G} = \frac{A_{UAC-1.3.3-G}}{B_{UAC-1.3.3-G}}. \quad (11)$$

The next step is to calculate the values of the quality subproperties.

The "Perceptiveness" subproperty is determined based on the quality attributes "Alternative Text", "Color contrast" and "Subtitles and audio descriptions" weighted by their respective weighting coefficients:

$$\begin{aligned}X_{UAC-1.1-G} &= w_{UAC-1.1.1-G} \cdot X_{UAC-1.1.1-G} + \\ &+ w_{UAC-1.1.2-G} \cdot X_{UAC-1.1.2-G} + \\ &+ w_{UAC-1.1.3-G} \cdot X_{UAC-1.1.3-G}.\end{aligned} \quad (12)$$

The "Operability" subproperty is determined based on the quality attributes "Keyboard navigation" and "Structured navigation", weighted by their respective weighting coefficients:

$$\begin{aligned}X_{UAC-1.2-G} &= w_{UAC-1.2.1-G} \cdot X_{UAC-1.2.1-G} + \\ &+ w_{UAC-1.2.2-G} \cdot X_{UAC-1.2.2-G}.\end{aligned} \quad (13)$$

The "Understandability" subproperty is determined based on the quality attributes "Clear instructions", "Input assistance" and "Correct input support" weighted by their respective weighting coefficients:

$$\begin{aligned}X_{UAC-1.3-G} &= w_{UAC-1.3.1-G} \cdot X_{UAC-1.3.1-G} + \\ &+ w_{UAC-1.3.2-G} \cdot X_{UAC-1.3.2-G} + \\ &+ w_{UAC-1.3.3-G} \cdot X_{UAC-1.3.3-G}.\end{aligned} \quad (14)$$

The "Localization" subproperty (UAC-2.1-S) is included as part of the "Supported languages adequacy" quality attribute. It measures the number of available language versions of the interface and is determined using the following formula:

$$X_{UAC-2.1-S} = w_1 \cdot A_{UAC-2.1-S} + w_2 \cdot B_{UAC-2.1-S} + w_3 \cdot C_{UAC-2.1-S} + w_4 \cdot D_{UAC-2.1-S}. \quad (15)$$

To evaluate "Localization" the weighting coefficient method is applied, which assigns significance to each language based on the linguistic environment of the country where the website was developed, as well as the resource's target audience in international markets.

In this study, the evaluation of the "Localization" subproperty is based on weighting coefficients specifically determined for a Ukrainian website. These coefficients take into account Ukraine's linguistic context, where Ukrainian is the primary language, English serves as an international communication medium, and German and French are among the most widely spoken languages in Europe. This approach ensures that the model is adapted to real-world conditions and meets the needs of the target audience. Thus, Ukrainian has the highest weighting coefficient (0.6) as it is the state language for the target audience. English is assigned a coefficient of 0.2 due to its importance as an international language of communication. German and French each have a coefficient of 0.08, reflecting their relevance for European users where these languages are widely spoken. Other languages receive the lowest coefficient (0.04), as they cover less significant audience segments.

The indicators $A_{UAC-2.1-S}$, $B_{UAC-2.1-S}$, $C_{UAC-2.1-S}$, $D_{UAC-2.1-S}$ take a value of 1 if the corresponding language is available on the website and 0 if it is absent.

The quality attribute values are derived from their respective subproperties, weighted by their corresponding weighting coefficients.

The "Accessibility for users with disabilities" attribute is determined using the following formula:

$$X_{UAC-1-G} = w_{UAC-1.1-G} \cdot X_{UAC-1.1-G} + w_{UAC-1.2-G} \cdot X_{UAC-1.2-G} + w_{UAC-1.3-G} \cdot X_{UAC-1.3-G}. \quad (16)$$

The "Supported languages adequacy" attribute is determined using the following formula:

$$X_{UAC-2-S} = w_{UAC-2.1-S} \cdot X_{UAC-2.1-S}. \quad (17)$$

The integral measure of the "Accessibility" quality subcharacteristic is determined as the sum of the products of each metric's value and its corresponding weighting coefficient:

$$X_{UAC} = w_{UAC-1-G} \cdot X_{UAC-1-G} + w_{UAC-2-S} \cdot X_{UAC-2-S}. \quad (18)$$

The integral indicator provides an assessment of the overall compliance of the university web portal with modern inclusivity requirements.

For a clearer representation of the structure of accessibility indicators, Figure 1 presents a classification model illustrating the distribution of key quality attributes, subproperties, and properties. This model visually highlights the essential aspects of web accessibility described earlier.

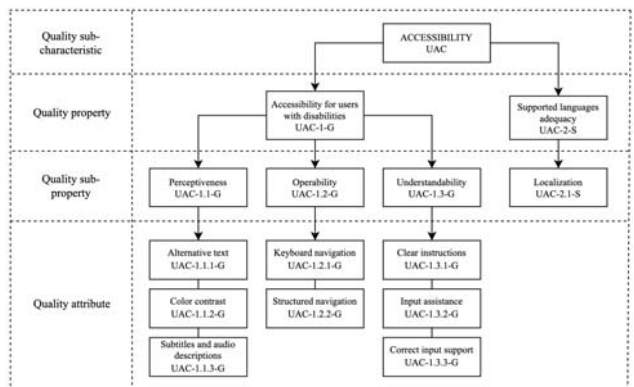

Figure 1 – Classification model of accessibility measures

A comprehensive approach to accessibility evaluation, covering perceptiveness, operability, understandability, and localization, enables a detailed analysis of all key aspects of user interaction with a web resource. The use of quantitative indicators and weighting coefficients ensures an objective assessment, allowing not only to determine the current level of compliance with accessibility standards but also to develop specific recommendations for improvement.

Thus, the proposed evaluation system not only reflects the overall level of accessibility but also helps identify the most problematic areas that require improvement. This contributes to enhancing the user experience for individuals with diverse needs, making the web portal more inclusive and user-friendly.

**4 EXPERIMENTS**

The methods developed in this study were applied to assess the accessibility of the website of Vasyl Stefanyk Precarpathian National University. University portals play a crucial role in providing access to information and services for a wide audience. Therefore, the website must be clear and user-friendly not only for students and faculty but also for prospective applicants, parents, international partners, and individuals with disabilities.

The accessibility of a web resource ensures equal access to educational materials, registration services, and general university information. Additionally, it enhances the institution's reputation, demonstrating its commitment to inclusivity and openness. In today's digital environ-ment, compliance with accessibility standards is not only a technical necessity but also a social imperative.

Next, the user interaction with key pages of the university website was evaluated.

Some accessibility metrics can be assessed more efficiently using automated tools and methodologies, which help streamline the evaluation process and provide objective results quickly [12]. For example, the Image Alt Checker service was used to analyze the alternative text of images (Figure 2), significantly reducing the workload involved in manual verification [13]. The color contrast assessment was conducted using the WCAG Contrast Checker extension (Figure 3), which automatically detects problematic elements on a page and ensures compliance with accessibility standards [14].

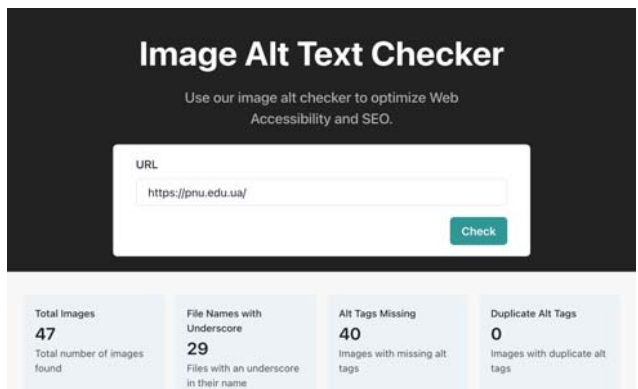

Figure 2 – Analysis results of the university website's homepage using the Image Alt Checker Service

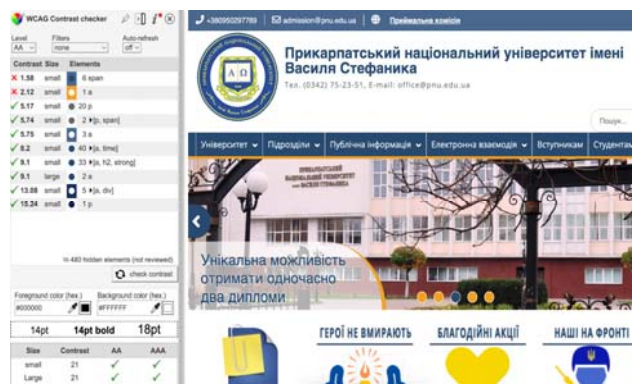

Figure 3 – Usage of the WCAG Contrast Checker plugin for color contrast evaluation

## 5 RESULTS

After calculating the values of each individual accessibility metric, their contribution to the overall assessment was determined through expert analysis. The expert evaluation method was used to establish weighting coefficients based on the importance of each quality indicator in shaping the overall accessibility level of the web resource.

To ensure clarity and ease of further calculations, the obtained values and weighting coefficients are presented in Table 1, allowing for a clear visualization of each metric's contribution to the final result.

Table 1 – Quality measures and their weights

| ID $j$ | Value $X_j$ | Weight $w_j$ |
|---|---|---|
| Quality attributes | | |
| UAC-1.1.1-G | 0.15 | 0.3 |
| UAC-1.1.2-G | 0.99 | 0.3 |
| UAC-1.1.3-G | 0 | 0.4 |
| UAC-1.2.1-G | 1 | 0.6 |
| UAC-1.2.2-G | 0.47 | 0.4 |
| UAC-1.3.1-G | 1 | 0.4 |
| UAC-1.3.2-G | 0 | 0.3 |
| UAC-1.3.3-G | 0.83 | 0.3 |
| Quality subproperties | | |
| UAC-1.1-G | 0.342 | 0.3 |
| UAC-1.2-G | 0.788 | 0.3 |
| UAC-1.3-G | 0.649 | 0.4 |
| UAC-2.1-S | 0.8 | 1.0 |
| Quality properties | | |
| UAC-1-G | 0.5986 | 0.6 |
| UAC-2-S | 0.8 | 0.4 |
| Quality subcharacteristic | | |
| UAC | 0.67916 | |

Based on the data presented in Table 1, several recommendations can be made to improve the accessibility indicators of the Vasyl Stefanyk Precarpathian National University website. Key areas for improvement include the "Subtitles and audio descriptions" and "Input assistance" attributes, which have zero values, indicating a lack of implementation and the need for significant enhancement. The "Alternative text" attribute has a low value of 0.15, suggesting that many images lack proper descriptions. The "Structured navigation" attribute has a value of 0.47, meaning improvements in page hierarchy and breadcrumb navigation would be beneficial. Several quality attributes already meet high accessibility standards, with values either equal to 1 or close to 1 (0.99 and 0.83), indicating no immediate need for refinement. Increasing the numerical values of the weaker quality attributes will, in turn, improve the values of subproperties, properties, and the overall "Accessibility" subcharacteristic of the website.

For a more precise analysis, it is necessary to interpret the results based on a defined scale that classifies the level of accessibility and establishes minimum acceptable compliance thresholds. The scale, shown in Figure 4, is used for this purpose. This scale is developed based on Fibonacci numbers (the "golden ratio") and enables the determination of the quality level of the evaluated software product. This scale ensures objectivity in assessment by allowing results to be classified as "very poor", "poor", "satisfactory", "good", or "excellent". It also defines the minimum acceptable values, which are crucial for determining whether the web resource meets modern accessibility standards.

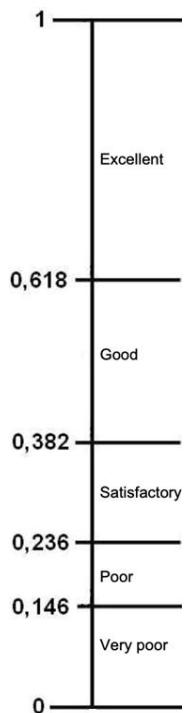

Figure 4 – Quality level scale for software products

According to the presented scale, the obtained overall accessibility assessment value of ≈0.68 falls within the "excellent" range, indicating a very high level of accessibility.

**6 DISCUSSION**

The developed mathematical model and accessibility evaluation methods demonstrate significant potential for analyzing and improving web resources in accordance with modern inclusivity standards.

Introducing new parameters into the subproperties of the "Accessibility" subcharacteristic allows for a more detailed assessment of this metric's quality.

The use of the proposed metrics enables a quantitative evaluation of accessibility at different stages of a software product's lifecycle, contributing to its improvement and enhancing the user experience.

The proposed methodology takes a systematic approach to analyzing web resources, focusing on key aspects of user interaction for individuals with disabilities.

Implementing the developed evaluation scale not only helps determine a website's compliance with modern accessibility requirements but also identifies specific areas that require improvement.

**CONCLUSIONS**

As a result of analyzing existing standardized accessibility evaluation metrics, it was found that while these metrics serve as a foundation for determining accessibility levels, they do not fully reflect real-world usage conditions and user needs due to their generalized approach. This highlights the necessity for further refinement and adaptation to enable more precise evaluation of specific digital products.

A classification and mathematical model was developed, and based on them, methods for evaluating software accessibility were designed.

The scientific novelty of this study lies in the development of evaluation methods for the "Accessibility" quality subcharacteristic by introducing new subproperties and quality attributes for software products. These are based on clearly defined metrics specifically adapted to assess the accessibility level of digital products for users with visual impairments. This approach ensures a more precise and objective assessment of web resources' compliance with inclusivity requirements, enhancing their effectiveness and usability for this user group.

An accessibility assessment was conducted for the main pages of the Vasyl Stefanyk Precarpathian National University website, allowing for an evaluation of its compliance with the expanded requirements of ISO 25023 [2]. The analysis covered key aspects such as content perceptiveness, interface operability, information clarity, and localization. The application of the proposed methodology not only enables an assessment of the current state of website accessibility but also provides practical recommendations for its improvement.

The practical significance of the obtained results lies in their application for objective accessibility evaluation of software products and web resources. This contributes to improving quality, ensuring compliance with international standards such as ISO 25023 [2] and WCAG [11], and promoting inclusivity, thereby expanding the user audience, including people with disabilities.

**ACKNOWLEDGMENTS**

The research was conducted as part of the scientific project "Development of a Methodology for Assessing the Quality of Software Products for Measuring Instruments" (state registration number 0116U002344).

УДК 004.05

## МЕТОДИ ОЦІНКИ ДОСТУПНОСТІ ПРОГРАМНИХ ПРОДУКТІВ

**Кузь М. В.** – д-р техн. наук, професор, професор кафедри інформаційних технологій, Прикарпатський національний університет імені Василя Стефаника, Івано-Франківськ, Україна.

**Яремій І. П.** – д-р фіз.-мат. наук, професор, професор кафедри матеріалознавства і новітніх технологій, Прикарпатський національний університет імені Василя Стефаника, Івано-Франківськ, Україна.

**Яремій Г. І.** – магістр, Прикарпатський національний університет імені Василя Стефаника, Івано-Франківськ, Україна.

**Пікуляк М. В.** – канд. техн. наук, доцент, доцент кафедри інформаційних технологій, Прикарпатський національний університет імені Василя Стефаника, Івано-Франківськ, Україна.

**Лазарович І. М.** – канд. техн. наук, доцент, доцент кафедри інформаційних технологій, Прикарпатський національний університет імені Василя Стефаника, Івано-Франківськ, Україна.

**Козленко М. І.** – канд. техн. наук, доцент, доцент кафедри інформаційних технологій, Прикарпатський національний університет імені Василя Стефаника, Івано-Франківськ, Україна.

**Векерик Д. В.** – магістр, Прикарпатський національний університет імені Василя Стефаника, Івано-Франківськ, Україна.

**АНОТАЦІЯ**

**Актуальність**. Розробка та вдосконалення методів оцінювання доступності програмних продуктів є актуальною задачею сучасної програмної інженерії, оскільки забезпечення рівного доступу до цифрових сервісів є ключовим фактором підвищення їхньої ефективності та інклюзивності. Зростаюча цифровізація суспільства вимагає створення програмного забезпечення, яке відповідає міжнародним стандартам доступності, таким як ISO/IEC 25023 та WCAG. Це дозволяє усувати бар'єри у використанні програмних продуктів людьми з різними фізичними, сенсорними та когнітивними потребами. Незважаючи на розвиток нормативних документів, існуючі методики оцінювання доступності часто мають узагальнений характер і не враховують специфічні потреби різних категорій користувачів, або особливості їх взаємодії з цифровими системами. Це створює необхідність розробки нових, більш деталізованих методів визначення показників, які впливають на якість взаємодії користувача із програмним продуктом.

**Мета**. Побудова класифікаційної та математичної моделі і розробка на її основі методів оцінювання доступності програмного забезпечення.

**Методи**. Розроблено метод оцінки підхарактеристики якості «Доступність», яка входить до складу характеристики якості «Зручність використання», що дало можливість виконати аналіз вебсайту на предмет інклюзивності для осіб із вадами зору та на його основі сформулювати конкретні рекомендації для подальшого вдосконалення, що є важливим кроком у напрямку створення інклюзивного цифрового середовища.

**Результати**. Запропоновано більш деталізовану та практично орієнтовану методику оцінювання доступності, у порівнянні із стандартизованими методиками. Використовуючи розроблену методику здійснено аналіз доступності основних сторінок вебсайту Прикарпатського національного університету імені Василя Стефаника та запропоновано вдосконалення вебсайту для підвищення його інклюзивності.

**Висновки.** У даному дослідженні виконано побудову класифікаційної та математичної моделі і розроблено методику оцінювання доступності вебсайтів на основі стандарту ISO 25023 та проведено аналіз основних сторінок вебпорталу університету. Визначені кількісні показники доступності дозволяють оцінити відповідність вебресурсу сучасним вимогам інклюзивності та сформувати рекомендації щодо його вдосконалення.

Наукова новизна полягає в розробці методів оцінки підхарактеристики якості «Доступність» шляхом введення нових підвластивостей та атрибутів якості програмних продуктів, що ґрунтуються на чітко визначених метриках, спеціально адаптованих для оцінювання рівня доступності цифрових продуктів для осіб із порушеннями зору. Такий підхід забезпечує більш точне та об'єктивне визначення відповідності вебресурсів вимогам інклюзивності, що сприяє підвищенню їхньої ефективності та зручності використання для зазначеної категорії користувачів.

Практичне значення отриманих результатів полягає в можливості їх застосування для об'єктивного оцінювання доступності програмних продуктів та веб-ресурсів.

**КЛЮЧОВІ СЛОВА:** доступність, інклюзивність, підвластивість якості, атрибут якості, перцептивність, керованість, зрозумілість, локалізація.